\documentclass[a4paper,12pt]{article}

\usepackage{amssymb}
\usepackage{amsmath}
\usepackage{mathbbol}
\usepackage{mathrsfs}
\usepackage{graphicx}
\usepackage{pgfplots}
\usetikzlibrary{patterns}
\usepackage{color}
\usepackage{xcolor}

\newtheorem{theorem}{Theorem}

\newtheorem{lemma}{Lemma}
\newtheorem{proposition}{Proposition}

\newcommand\unit{\hbox{\rm 1\kern-2.8truept l}}

\newcommand\mi{\mathrm{i}}
\newcommand{\tr}[1]{\hbox{\rm{tr}} (#1)}
\newcommand\Dom{\hbox{\rm Dom}}

\newcommand\Lform{{\mathcal{L}}\kern-9pt\raise1.0pt\hbox{$-$}}
\newcommand{\h}{\mathsf{h}}

\colorlet{gold}{red!10!yellow!90!}
\colorlet{Franco}{blue!50!black!50!}

\begin{document}

\title{Quadratic Open Quantum Harmonic Oscillator}

\date{}





\author{}

\maketitle

\begin{center}
Ameur Dhahri\footnote{Dipartimento di Matematica, Politecnico di Milano,
Piazza Leonardo da Vinci 32, I-20133 Milano (Italy), {\tt ameur.dhahri@polimi.it}},
Franco Fagnola\footnote{Dipartimento di Matematica, Politecnico di Milano,
Piazza Leonardo da Vinci 32, I-20133 Milano (Italy) {\tt franco.fagnola@polimi.it}} and
Hyun Jae Yoo\footnote{Department of Applied Mathematics, Hankyong National University,
327 Jungang-ro, Anseong-si, Gyeonggi-do 456-749, Korea
{\tt yoohj@hknu.ac.kr}}
\end{center}

\begin{abstract}
We study the quantum open system evolution described by a
Gorini-Kossakowski-Sudarshan-Lindblad generator with creation and annihilation operators
arising in Fock representations of the $\mathfrak{sl}_2$ Lie algebra. We show that
any initial density matrix evolves to a fully supported density matrix and converges
towards a unique equilibrium state. We show that the convergence is exponentially fast
and we exactly compute the rate for a wide range of parameters.
We also discuss the connection with the two-photon absorption and emission process.
\end{abstract}

\noindent{Keywords: quantum harmonic oscillator, quantum Markov semigroup, Fock representations
of the $\mathfrak{sl}_2$ algebra, spectral gap.}

\noindent{Subject Classification: 46L55, 82C10, 60J27.}

\section{Introduction}

Models of quantum harmonic oscillators are usually based on commutation relations.
The Heisenberg-Weyl algebra commutation relations $[H,A]=-A$, $[H,A^+]=A^+$,
$[A,A^+]=\mathrm{1}$, or in terms of position $Q=(A^+ + A)/\sqrt{2}$ and momentum
$P=\mathrm{i}(A^+-A)/\sqrt{2}$, $[H,Q]=-\mathrm{i}P$, $[H,P]=\mathrm{i}Q$,
$[Q,P]=\mathrm{i}$ are the foundation at the best known one. This model
arises, for instance, replacing time derivatives in the classical equation $q'' = -q$
by commutators with the Hamiltonian operator $H=(P^2+Q^2)/2$ so that we can write
it as $[H,[H,Q]]=Q$. If we fix $H$ and define $P=\mathrm{i}[H,Q]$, then the double
commutator equation reads as $[H,P]=\mathrm{i} Q$ and if, moreover, we want $H,P,Q$
to be elements of a Lie algebra, the Jacobi identity $[H,[P,Q]]+[P,[Q,H]]+[Q,[H,P]]=0$
implies that $[P,Q]$ commutes with $H$. The most natural choice as $-\mathrm{i}\mathbb{1}$
corresponds to the commutation relations of the Heisenberg-Weyl algebra.

Other choices lead to different models of quantum oscillators (see, for instance, \cite{AtAt}
and the references therein) and for some of them it is possible to develop a complete theory
describing explicitly spectra of observables, eigenvectors, time evolution, etc.
The choice $[P,Q]= -2\mathrm{i}H$ corresponds to the commutation relations of the
$\mathfrak{sl}_2$ Lie algebra.

This is a three dimensional simple $^*$-Lie algebra with basis $\{B^+,B,M\}$, commutation
relations $[B,B^+]=M$, $[M,B^+]=2B^+$, $[M,B]=-2B$ and involution $B^*=B^+$, $M^*=M$.
The construction of Fock representations of $\mathfrak{sl}_2$ Lie algebra and of the
current algebra associated to its central extension motivated a large number of papers
extending it in different directions: see Ref.\cite{Snia00} for the case of free white
noise; \cite{AcFrSk} for the connection with quantum L\'evy processes;
Refs.\cite{AcDh09,AcDh,D1,D2} for the construction of the quadratic
Fock functor.

The weak coupling limit (see \cite{AcLuVo}) of an harmonic oscillator coupled with a
reservoir in equilibrium with inverse temperature $\beta>0$ gives rise
to a fundamental model of an open quantum system with
a lot of deep properties and quantities that can be computed explicitly
called in the literature the {\it open quantum harmonic oscillator}
(see e.g. Ref.\cite{IsSc} and the references therein).
If we consider, instead, the formal Gorini-Kossakowski-Sudarshan-Lindblad
(GKSL) generator arising in the weak coupling
limit of an oscillator based on the Fock representations of $\mathfrak{sl}_2$ commutation
relations we find
\begin{eqnarray}\label{PHE}
\mathcal{L}(x) & = &-\frac{\lambda^2}{2}\left( B B^+ x - 2 B x B^+ + x B B^+ \right) \\
& & -\frac{\mu^2}{2}\left( B^+ B x - 2 B^+ x B  + x B^+ B \right)
+\mi [\zeta^+ B B^+ + \zeta^{-} B^+ B, x\,]\nonumber
\end{eqnarray}
where $\zeta^\pm,\lambda,\mu$ are real parameters and $\lambda,\mu>0$.
This is called the {\it quadratic open quantum harmonic oscillator} because the operators
$B,B^+$ are the annihilation and creation operators arising in Fock representations of
$\mathfrak{sl}_2$ and the action of $BB^+$ and $B^+B$ (see formulae \eqref{eq:BB+}) is quadratic
with respect to the level of the system while, for the open quantum harmonic oscillator,
it is linear. Constants $\lambda^2,\mu^2$ are related with the inverse
temperature $\beta$ by $\lambda^2\mu^{-2}=\hbox{\rm e}^{-s\beta}$ for some $s>0$.

This is a simple and natural model, however, contrary to what happens for the open
quantum harmonic oscillator, it does not admit explicit solutions except for the
formula of the invariant state. As an example, if one looks at the action on the abelian
algebra of functions of the number operator, one finds a birth-and-death process
with quadratic jump rates for which explicit representations for transition
probabilities, to the best of our knowledge (see Ref.\cite{SiVa}), are not known.

In this paper, we first show that the formal GKSL generator
with unbounded operators $B,B^+$ generates a unique quantum Markov semigroup
and we establish the existence of a unique explicit equilibrium state.
Then, we study the behaviour of the evolution
of states and observables for all values of parameters involved. We prove that
any initial state converges towards the unique equilibrium state for the
trace norm (Theorem \ref{th:trace-norm-conv}). We also prove (Theorem \ref{th:inst-spread})
that any initial state $\rho_0$, in particular also a pure state, evolves to a faithful
state $\rho_t$ for all $t>0$. Moreover, we show that, for some special
values of a parameter $r$ determining the Fock representation of the $\mathfrak{sl}_2$
commutation relations this model is intimately related with the two-photon
absorption and emission process studied in \cite{CaFaGaQu,FaQu}.
Finally, we show that convergence towards the unique invariant
state is exponentially fast (with respect to the Hilbert-Schmidt norm induced by the
invariant state) and we also compute the sharp exponential rate for a wide range of parameters
(Theorem \ref{thm:gap}). Our analysis shows, in particular, that the decay rate of
off-diagonal terms of density matrices is smaller than the rate of convergence of
the diagonal part towards the unique equilibrium state for more and more values
of the parameter $r$ as the inverse temperature $\beta$ becomes big, i.e. the
reservoir becomes cooler. In other words, at low temperatures, decoherence is
slower than relaxation for $r$ away from $0$.

The paper is organized as follows. In Section \ref{sect-model}, we introduce the model
of the quadratic open quantum harmonic oscillator. The full characterization of invariant
states and the asymptotic behaviour of the associated quantum Markov semigroup are studied
in Section \ref{sect:inv-stat}. The close relationship with two-photon
absorption and emission process is studied in Section \ref{sect:2photon}.
In Section \ref{sect:inst-spread}, we show that for all initial state the support
of the state evolved at any time $t>0$ is full.
The rate of the exponentially fast convergence towards the unique invariant
state is studied in Section \ref{sect:gap}.

\section{The model}\label{sect-model}
Let $\h$ be the Hilbert space $\h=\ell^2({\mathbb{N}})\simeq\Gamma(\mathbb{C})$
with canonical orthonormal basis $(e_n)_{n\ge 0}$.
We consider the operators $B, B^+, M$ of the Fock representation
of the renormalized square of the white noise Lie algebra
$B$, $B^+$ and $M$ with domain
\[
\Dom(B) =\Dom(B^+)= \Dom(M) =
\left\{ u = \sum_{n\ge 0} u_n e_n \,\Big|\, \sum_{n\ge 0} n^2|u_n|^2 <\infty\right\}
\]
defined, on vectors of the canonical orthonormal basis, by
\begin{eqnarray}\label{eq:BB+}
B e_{n} & = & \omega_n^{1/2} e_{n-1}, \ \hbox{for} \ n>0, \qquad Be_0=0,
\nonumber \\
B^+ e_n & = & \omega_{n+1}^{1/2} e_{n+1} \\
M e_n & = & (2n+r) e_n, \nonumber
\end{eqnarray}
where $r > 0$ is a real parameter (see Ref.\cite{AcFrSk}
Section 3.2 p. 134 for the explanation why $r$ must be non-negative) and
\[
\omega_n = n(n+r-1)
\]
which is strictly positive for all $n\ge 1$ and satisfies $\omega_0=0$.

Note that the domain of $B,B^+$ and $M$ coincides with the domain of the
number operator $N$ defined by $Ne_n=ne_n$ for all $n\ge 0$.

We consider the formal Lindblad generator (\ref{PHE})
which is of weak coupling limit type (see Refs.\cite{AcFaQu,AcLuVo})
since it arises in the weak coupling limit of a system with Hilbert space
$\ell^2(\mathbb{N})$ and Hamiltonian $H_S$ given by the number
operator $N$ coupled to a Boson reservoir in equilibrium
with inverse temperature $\beta>0$ and interaction operator
\[
B\otimes A^+(g) + B^+ \otimes A(g).
\]
Constants $\lambda^2,\mu^2$ satisfy $\lambda^2\mu^{-2}
=\hbox{\rm e}^{-s\beta}$ for some $s>0$.

Moreover (see Section \ref{sect:2photon}) for $r=1/2$ (resp. $r=3/2$) and a suitable
choice of the real constants $\zeta^{-},\zeta^{+}$ we find the even (resp. odd)
part of the two-photon absorption and emission generator studied in
Refs.\cite{CaFaGaQu,FaQu}.

Let $G$ be the operator defined on the domain
${\rm Dom}(N)$ of the number operator by
\[
G=-\frac{\lambda^2}{2}B B^+ -\frac{\mu^2}{2}B^+B
- \mi \left( \zeta^+ B B^+ + \zeta^{-} B^+ B\right)
\]
and let $L_1,L_2$ be the operators defined on ${\rm Dom}(N)$ by
\[
L_1=\mu B,\qquad L_2 = \lambda B^+.
\]
Clearly $G$ is a function of the number operator $N$,
defined on the same domain of $B,B^+$ and $M$, since
\[
G = -\left(\frac{\lambda^2}{2}+\mi\zeta^{+}\right)\omega_{N+1}
- \left(\frac{\mu^2}{2}+\mi\zeta^{-}\right)\omega_{N}
\]
with negative real part hence generates a strongly continuous semigroup
of contractions $(P_t)_{t\geq 0}$ on $\h$ explicitly given by
\[
P_t e_n = \hbox{\rm e}^{-t\left(\left(\frac{\lambda^2}{2}+\mi\zeta^{+}\right)\omega_{n+1}
+ \left(\frac{\mu^2}{2}+\mi\zeta^{-}\right)\omega_{n}\right)} e_n.
\]
For every $x\in{\mathcal B}(\h)$ the formal generator is the sesquilinear form
\begin{equation}\label{eq:Lform-quadratic}
{\Lform}(x)[u,v]=\langle Gu, xv\rangle + \sum_{\ell=1}^{2}\langle
L_{\ell}u, xL_{\ell}v\rangle + \langle u, xGv\rangle,
\end{equation}
for $u, v\in {\rm Dom}(G)={\rm Dom}(N^2)$. One can easily check
that conditions for constructing the minimal quantum dynamical
semigroup (QDS) associated with the above $G, L_1, L_2$ ((H-min)
in Ref.\cite{ffproy}) hold and this semigroup ${\mathcal T}=
({\mathcal T}_t)_{t\geq 0}$ satisfies the so-called Lindblad
equation
\begin{equation}\label{eq:minQMS}
\langle v, {\mathcal T}_t(x)u\rangle= \langle v, P_{t}^*x
P_{t}u\rangle +\sum_{\ell=1}^{2}\int_{0}^{t}\langle
L_{\ell}P_{t-s}v, {\mathcal T}_{s}(x)L_{\ell}P_{t-s}u\rangle\,ds,
\end{equation}
for all $u, v\in {\rm Dom}(G)$.

A straightforward computation using the CCR (it could be done considering
quadratic forms on the linear manifold generated by vectors $(e_{n})_{n\ge 0}$
if one wants to cope with unboundedness of the involved operators but we
prefer to simplify the notation) shows that
\begin{eqnarray*}
\Lform(f(N)) & = & \lambda^2 (N+1)(N+r)\left(f(N+1)-f(N)\right) \\
& + & \mu^2 N(N+r-1)\left( f(N-1)-f(N)\right).
\end{eqnarray*}
Taking $f(n)=(n+1)^2$, for $n \ge 1-r$ we easily find
\begin{eqnarray*}
& & \lambda^2 (n+1)(n+r)\left(f(n+1)-f(n)\right)
 +  \mu^2 n(n+r-1)\left( f(n-1)-f(n)\right) \\
& = & 2(\lambda^2-\mu^2)n^3 + (\lambda^2(2r+3)+\mu^2(2r-3))) n^2 \\
& + & (\lambda^2 (3r+1)+\mu^2(r-1))n + \lambda^2 r
\end{eqnarray*}
and, for $0\le n < 1-r$, i.e. $n=0$ we obviously find $0$.
Therefore, defining as $b$ the maximum of the three constants
\[
\left|\lambda^2(2r+3)+\mu^2(2r-3)\right|, \
\frac{\left|\lambda^2 (3r+1)+\mu^2(r-1)\right|}{2}, \
\lambda^2 |r|,
\]
we have
\[
\Lform ((N+\unit)^2) \le - 2(\mu^2-\lambda^2) N^3 + b (N+\unit)^2.
\]

As a consequence, if $\lambda\leq\mu$, $\Lform$ satisfies a well known criterion
for conservativity (Ref.\cite{ffproy} Theorem 3.40). Moreover, for $\lambda >\mu$
the formal generator satisfies a simple criterion for nonconservativity,
see Ref.\cite{garcia-quezada}, Example 2. Then the minimal QDS is Markov
(or conservative) if and only if $\lambda\leq\mu$.
It follows from conservativity that the minimal QDS is the unique solution of
equation (\ref{eq:minQMS}).
Moreover an operator $x\in{\mathcal{B}}(\h)$ belongs to the domain of the
generator ${\mathcal{L}}$ if and only if the sesquilinear form $\Lform(x)$ is
bounded (see Ref.\cite{ffproy} Prop. 3.33 p.64).

The action of ${\mathcal L}$ on the linear manifold
$ {\cal M}={\rm span}\{|e_j\rangle\langle e_k|:j,k\geq 0\}$ of finite range operators is given by
\begin{eqnarray}
 {\mathcal{L}}(x)
& = &  \mi\sum_{j,k}(\zeta^{+}(\omega_{j+1}-\omega_{k+1})
+\zeta^{-}(\omega_j-\omega_k))x_{jk} |e_j\rangle\langle e_k| \label{eq-Lindblad-on-M}\\
& + &  \sum_{j,k}\Big(\mu^2\omega_k^{1/2}\omega_j^{1/2}x_{j-1\,k-1}-
{\frac{\mu^2}{2}}(\omega_j+\omega_k)x_{jk} \nonumber \\
&+&  \lambda^2\omega_{j+1}^{1/2}\omega_{k+1}^{1/2}x_{j+1\, k+1}
-{\frac{\lambda^2}{2}}(\omega_{j+1}+\omega_{k+1})x_{jk}\Big)|e_j\rangle\langle e_k|.\nonumber
\end{eqnarray}

\section{Invariant states and asymptotic behaviour}
\label{sect:inv-stat}

The behaviour of the quadratic open quantum harmonic oscillator and the
structure of its invariant states depends crucially upon the parameters
$\lambda$ and $\mu$. We begin by considering the case where
$\mu>\lambda>0$.

\begin{proposition}\label{prop:inv-state}
If $\nu=\lambda/\mu < 1$ then the normal state
\begin{equation}\label{eq:inv-state}
\rho = (1-\nu^2)\sum_{n\ge 0} \nu^{2n} |e_n\rangle\langle e_n|
\end{equation}
is invariant.
\end{proposition}

\noindent{\bf Proof.}
Let ${\mathcal L}_{*}$ be the generator of the predual semigroup ${\mathcal T}_{*}
=({\mathcal T}_{*t})_{t\geq 0}$, acting on the Banach space  of trace
class operators on $\h$. Consider the approximations
${\rho}_{n}=(1-\nu^2)\sum_{k=0}^{2n}\nu^{2k} |e_k\rangle\langle e_k|$,
of $\rho$ by finite rank operators.

The operators ${\rho}_{n}$ belong to the domain of ${\mathcal L}_{*}$ and
we can write ${\mathcal L}_{*}(\rho_{n})$ as $(1-\nu^2)$ times
\begin{eqnarray*}
&  & \kern-20truept \sum_{k=0}^n \lambda^2 \nu^{2k}\omega_{k+1}
\left(|e_{k+1}\rangle\langle e_{k+1}|-|e_k\rangle\langle e_k|\right)
 +\sum_{k=1}^n\mu^2\nu^{2k}\omega_k
 \left(|e_{k-1}\rangle\langle e_{k-1}|-|e_k\rangle\langle e_k|\right)  \\
& = & \sum_{k=1}^{n}\left(\lambda^2\nu^{2(k-1)}-\mu^2\nu^{2k}\right)\omega_k|e_k\rangle\langle e_k|
+ \sum_{k=0}^{n-1}\left(\mu^2\nu^{2(k+1)}-\lambda^2\nu^{2k}\right)|e_k\rangle\langle e_k| \\
& & + \lambda^2 \nu^{2n} \left(\omega_{n+1}|e_{n+1}\rangle\langle e_{n+1}|
-\omega_n |e_n\rangle\langle e_n|\right).
\end{eqnarray*}
Terms in the above summations vanish because $\mu^2\nu^{2(k+1)}=\lambda^2\nu^{2k}$
for all $k\ge 0$. Moreover
\[
\lim_{n\to\infty}\left\Vert \mathcal{L}_*(\rho_n)\right\Vert_1
= \lim_{n\to\infty} \lambda^2 \nu^{2n} (\omega_{n+1}+\omega_n) =0
\]
because $\lambda < \mu$. Since the operator $\mathcal{L}_*$ is closed,
it follows that $\rho$ belongs to the domain of $\mathcal{L}_*$ and
$\mathcal{L}_*(\rho)=0$.

\medskip

In order to show uniqueness of the invariant state
(\ref{eq:inv-state}) we begin by recalling that the support projection
$p$ of an invariant state with density matrix $\rho$, i.e. the orthogonal
projection onto the range of $\rho$, satisfies $\mathcal{T}_t(p)\ge p$
for all $t\ge 0$ (see e.g. \cite{FagReb-subharm-proj} Theorem II.1).
Such projections, called subharmonic, are easily characterized in terms
of invariant subspaces of operators $P_t$ and $L_1,L_2$ considered in
Section \ref{sect-model}. A QMS is called \emph{irreducible} if the only
subharmonic projections are the trivial ones $0,\unit$. In this case,
it is well-known (see Ref.\cite{FrLMP77} Lemma 1) that a faithful invariant state,
if it exists, is unique because the set of fixed points for the QMS $\mathcal{T}$
is the trivial algebra $\mathbb{C}\unit$. In our framework we can prove the following.

\begin{proposition}\label{prop:irred}
The QMS $\mathcal{T}$ is irreducible for all $\lambda\le \mu$. In particular, if $\lambda < \mu$,
the state (\ref{eq:inv-state}) is the unique $\mathcal{T}$-invariant state.
\end{proposition}

\noindent{\bf Proof.}
The range of any non-trivial subharmonic projection determines an
invariant subspace for the operators $P_t$ for all $t>0$ (see
Ref.\cite{FagReb-subharm-proj} Theorem III.1). Since these
operators are normal and compact, these invariant subspaces are generated
by eigenvectors of $P_t$. Moreover, knowing the spectral decomposition
of $P_t$ (it is a function of the number operator!) we infer that
they are generated by collections of vectors $(e_n)_{n\in I}$ for
some subset $I$ of $\mathbb{N}$. Invariance of these subspaces for
$B$ and $B^+$ implies then that they must coincide with the
whole of $\h$. This proves that the QMS is irreducible.

If $\lambda < \mu$ the QMS admits the faithful invariant state (\ref{eq:inv-state})
and so the set of fixed points for the QMS $\mathcal{T}$ is the trivial
algebra $\mathbb{C}\unit$. It follows then from Lemma 1 of Ref.\cite{FrLMP77}
that (\ref{eq:inv-state}) is the unique invariant state.
\hfill $\square$

\smallskip
Applying the main result of Ref.\cite{DhFaRe} we can also
show convergence towards the invariant state in trace norm.
As a preliminary step we prove the following
result which is interesting on its own

\begin{proposition}\label{prop:FT-NT-trivial}
If $\nu=\lambda/\mu < 1$  the decoherence free subalgebra
\[
\mathcal{N}(\mathcal{T})= \left\{ x\in\mathcal{B}(\h) \mid
              \mathcal{T}_t(x^*x)=\mathcal{T}_t(x^*)\mathcal{T}_t(x), \
              \mathcal{T}_t(x x^*)=\mathcal{T}_t(x)\mathcal{T}_t(x^*), \, \forall t\ge 0
                 \right\}.
\]
and the fixed point algebra
$ \mathcal{F}(\mathcal{T}) = \left\{ x\in\mathcal{B}(\h) \mid
              \mathcal{T}_t(x) = x \ \forall t\ge 0 \right\}$
are trivial.
\end{proposition}

\noindent{\bf Proof.}
It is well-known that $\mathcal{N}(\mathcal{T})$ is a von Neumann subalgebra
of $\mathcal{B}(\h)$  (see e.g. Proposition 2.1 (3) of Ref.\cite{DhFaRe}).
Moreover, since the invariant state $\rho$ defined in (\ref{eq:inv-state}) is
faithful, also $\mathcal{F}(\mathcal{T})$ is a von Neumann subalgebra of
$\mathcal{B}(\h)$. Indeed, if $x$ belongs to ${\mathcal{F}}(\mathcal{T})$, then,
by $2$-positivity, $\mathcal{T}_t(x^*x)\ge \mathcal{T}_t(x^*)\mathcal{T}_t(x)=x^*x$
and $\tr{\rho(\mathcal{T}_t(x^*x)-x^*x)}=0$ because $\rho$ is invariant.
It follows that $\mathcal{T}_t(x^*x)=x^*x$ i.e. $x^*x\in \mathcal{F}(\mathcal{T})$.

As a by product, if $x\in\mathcal{F}(\mathcal{T})$, then
\[
\mathcal{T}_t(x^*x)= x^* x = \mathcal{T}_t(x^*)\mathcal{T}_t(x),
\]
and the same identity holds exchanging $x$ and $x^*$,
i.e. $\mathcal{F}(\mathcal{T})$ is contained in $\mathcal{N}(\mathcal{T})$.

Thus, it suffices to prove that $\mathcal{N}(\mathcal{T})$ is trivial.
To this end, we apply Theorem 4.1 of Ref.\cite{DhFaRe} characterizing
$\mathcal{N}(\mathcal{T})$ as the generalized commutator of the set
of unbounded operators
\begin{equation}\label{eq-gen-comm}
{\mathcal{D}}\left(\mathcal{T}\right)
:=\left\{\, \hbox{\rm e}^{-\mi tH} L_\ell\, \hbox{\rm e}^{\mi tH}\, ,\,
\hbox{\rm e}^{-\mi tH} L_\ell^*\, \hbox{\rm e}^{\mi tH}\,
\mid\, \ell\ge 1,t\ge 0\right\}.
\end{equation}
where $H=\zeta^+ B B^+ + \zeta^{-} B^+ B$.
The additional technical domain assumptions that can be
easily checked taking as $D$ the linear manifold spanned by
finite linear combinations of vectors $e_n$ of the orthonormal basis
and as operator $C$ the number operator or $(N+\mathbb{1})^2$.

If $X$ is an operator in the generalized commutator of (\ref{eq-gen-comm}),
then it is, by definition of generalized commutator, bounded and, in particular,
it satisfies
\[
XB \subseteq BX, \qquad XB^+ \subseteq B^+X
\]
(meaning that $BX$ is an ampliation of $XB$ and $B^+X$ is an ampliation of $XB^+$).
It follows that
\[
XB^+B \subseteq B^+XB \subseteq B^+BX, \qquad  XBB^+ \subseteq BXB^+ \subseteq BB^+X,
\]
and so, since the difference $B B^+ - B^+ B$ is $2N+r\unit$, $NX$ is an ampliation
of $XN$ and
\[
X(s+N)\subseteq (s+N)X
\]
for all $s>0$. Left and right multiplying by the resolvent $(s+N)^{-1}$,
since the operators $(s+N)^{-1} X$ and $X(s+N)^{-1}$ are bounded, we find
$(s+N)^{-1}X=X(s+N)^{-1}$ for all $s>0$. This shows that $X$ commutes with
every spectral projection $|e_n\rangle\langle e_n|$ of the number operator
and so it is a function $f(N)$ of the number operator itself.
However, since $\omega_{n+1}>0$ for all $n\ge 0$ if $r>0$
\[
f(N)B^+ e_n - B^+ f(N) e_n = \omega_{n+1}^{1/2}(f(n+1)-f(n))e_n
\]
vanishes if and only if $f$ is constant and so the generalized commutator
of (\ref{eq-gen-comm}) is trivial. \hfill $\square$

\medskip

We are now in a position to prove the following.

\begin{theorem}\label{th:trace-norm-conv}
If $0<\nu=\lambda/\mu < 1$
then (\ref{eq:inv-state}) is the unique invariant state and
\[
\lim_{t\to \infty}\mathcal{T}_{*t}(\eta) = \rho
\]
in trace norm for all initial state $\eta$.
\end{theorem}

\noindent{\bf Proof.}
Since $\mathcal{F}(\mathcal{T})=\mathcal{N}(\mathcal{T})
=\mathbb{C}\mathbb{1}$ by Proposition \ref{prop:FT-NT-trivial},
the conclusion is immediate from Theorem 3.3 of Ref.\cite{DhFaRe}.
\hfill $\square$

\medskip

We complete the study of the asymptotic behaviour by the following.
\begin{proposition} \label{prop:transient}
If $\lambda=\mu>0$ the QMS is transient. In particular, it has
no invariant state.
\end{proposition}
\noindent{\bf Proof.}
By Theorem 5 Ref. \cite{FFRR03}, it suffices to find a strictly positive bounded
operator $X$ such that $\mathcal{T}_t(X)\le X$ for all $t\ge 0$. Inspired by a
result on classical birth and death processes (\cite{Karl-McGreg} Theorem 2
and Lemma 1), we consider the operator
\[
X = \left(\sum_{k\ge n} (k+1)^{-1}(k+r)^{-1}\right)|e_n\rangle\langle e_n|
\]
which is clearly bounded since $\sum_{k\ge 0} (k+1)^{-1}(k+r)^{-1} <\infty$
and is a function of the number operator. A straightforward computation shows
that
\[
\Lform(X) = - |e_0\rangle\langle e_0| \le 0.
\]
It follows that $X$ belongs to the domain of $\mathcal{L}$  (\cite{ffproy} Prop. 3.33 p.64)
and
\[
\frac{d}{dt}\mathcal{T}_t(X) = -\mathcal{T}_t(|e_0\rangle\langle e_0|) \le 0
\]
so that $\mathcal{T}_t(X)\le X$ for all $t\ge 0$.
Since the QMS $\mathcal{T}$ is transient, it has no invariant state by Proposition
6 of Ref.\cite{FFRR03}.
\hfill $\square$

\medskip

In the case where there is a faithful invariant state, it is not difficult
to show that the quantum detailed balance condition (in most of its
quantum formulations as in \cite{FFVU10}) holds.

\section{Relationship with the two-photon absorption and emission process}
\label{sect:2photon}

The two-photon absorption and emission process quantum Markov semigroup is
generated by
\begin{eqnarray*}
\mathcal{L}_{\rm tp}(x) & = & \mi\, [\xi^{-}a^{+2} a^2+\xi^+ a^2 a^{+2}, x]  \\
& - & \frac{\lambda^2}{2}\left( a^2a^{+2} x - 2 a^2 x a^{+2} + x a^2 a^{+2}\right) \\
& - &\frac{\mu^2}{2}\left( a^{+2}a^2 x - 2 a^{+2} x a^2 + x a^{+2} a^2\right)
\end{eqnarray*}
where $a,a^+$ are the usual annihilation and creation operators in
$\h$, $\lambda,\mu,\xi^{+},\xi^{-}\in\mathbb{R}$, $\lambda\le \mu$.

This generator has been studied in Ref.\cite{FaQu} for $\xi^{+}=0$, however,
this does not change any result of that paper. In particular, it has been proved
that the even and odd projections
\[
p_{\rm e} = \sum_{k\ge 0} |e_{2k}\rangle\langle e_{2k}|, \qquad
p_{\rm o} = \sum_{k\ge 0} |e_{2k+1}\rangle\langle e_{2k+1}|
\]
are harmonic (i.e. invariant) for the QMS generated by $\mathcal{L}_{\rm tp}$.
As a consequence we can consider the restricted semigroups $\mathcal{T}^{\rm e}$
and $\mathcal{T}^{\rm o}$ on the von Neumann subalgebras
$p_{\rm e}\mathcal{B}(\h)p_{\rm e}$ and  $p_{\rm o}\mathcal{B}(\h)p_{\rm o}$,
identified respectively with $\mathcal{B}(p_{\rm e}\h)$ and
$\mathcal{B}(p_{\rm o}\h)$.

Let $U_{\rm e}, U_{\rm o}$ be the unitary operators
\[
U_{\rm e}:\h\to p_{\rm e}\h, \quad
U_{\rm e} e_{k} = e_{2k}, \qquad
U_{\rm o}:\h\to p_{\rm o}\h, \quad
U_{\rm o} e_{k} = e_{2k+1}
\]
A straightforward computation shows that, if $r=1/2$,
\[
U_{\rm e} B U^*_{\rm e} e_{2k} = 2\, a^2 e_{2k}, \qquad
U_{\rm e} B^+ U^*_{\rm e} e_{2k} = 2\, a^{+2} e_{2k}
\]
and, if $r=3/2$, similarly
\[
U_{\rm o} B U^*_{\rm o} e_{2k+1} = 2\, a^2 e_{2k+1}, \qquad
U_{\rm o} B^+ U^*_{\rm o} e_{2k+1} = 2\, a^{+2} e_{2k+1},
\]
so that, in both cases,
\begin{eqnarray*}
U_{\rm e} \left(\zeta^{+}B B^+ +\zeta^{-} B^+ B\right)U^*_{\rm e}
= 4 \left(\zeta^{+} a^2 a^{+2} + \zeta^{-} a^{+2} a^2 \right)
\end{eqnarray*}
As a consequence, the quadratic open quantum harmonic oscillator generator
$\mathcal{L}$, for $\xi^\pm = 4\zeta^\pm$, satisfies
\begin{eqnarray*}
U_{\rm e}\mathcal{L}(U^*_{\rm e} x U_{\rm e}) U^*_{\rm e}
= 4 \,\mathcal{L}_{\rm tp}(p_{\rm e} x p_{\rm e})
& \qquad & \hbox{\rm for } \ r=1/2 \\
U_{\rm o}\mathcal{L}(U^*_{\rm o} x U_{\rm o}) U^*_{\rm o}
= 4\, \mathcal{L}_{\rm tp}(p_{\rm o} x p_{\rm o})
& \qquad & \hbox{\rm for } \ r=3/2
\end{eqnarray*}

This shows that the QMS of the quadratic open quantum
harmonic oscillator is unitarily equivalent (up to the multiplicative constant
$4$) to the QMS obtained by restriction of the two-photon absorption and
emission process to the even (resp. odd) states of the number operator
for $r=1/2$ (resp. $r=3/2$), for a suitable choice of the parameters
$\xi^\pm,\zeta^\pm$. This analogy will serve as an inspiration to study
the exponential speed of convergence towards the equilibrium state.

\section{Instantaneous spread of state supports}\label{sect:inst-spread}

In this section we will show that for all initial state $\eta$ the support of the
state $\mathcal{T}_{*t}(\eta)$ at any time $t>0$ is the whole of $\h$.

This property follows from a recent result by Hachicha, Nasroui
and Gliouez \cite{HaNaGl} Theorem 3.3 for QMSs associated with operators $G,L_\ell$
(in our case $L_1=\lambda B^+, L_2=\mu B$) constructed form generators.

\begin{theorem}\label{th:HaNaGl}
Suppose that $G$ generates an analytic semigroup in a sector
$\Delta=\lbrace z \in \mathbb{C} \mid | \hbox{\rm arg } z | < \theta \rbrace$
with $\theta \in ]0,\frac{\pi}{2}]$ and, moreover, that
\begin{equation}\label{eq:LGk}
L_{\ell}(\Dom(G^{k})) \subseteq \Dom(G^{k-1})
\end{equation}
for all $k> 0.$
For all state $\omega=\sum_{j \in J}\omega_{j}|\psi_{j}\rangle \langle \psi_{j}|,$
with $\omega_{j}> 0$ for all $j\in J$ and all $t \geq 0,$ the support projection
$\mathcal{S}_{t}(\omega)$ of the state $\mathcal{T}_{*t}(\omega)$
is the closure of linear manifold generated by vectors
\begin{equation}\label{eq:support}
P_{t}\psi_j, \
 \delta_{G}^{m_{1}}(L_{\ell_1})\delta_{G}^{m_{2}}(L_{\ell_2})\cdots
 \delta_{G}^{m_{n}}(L_{\ell_n})P_{t}\psi_{j}
\end{equation}
for all $j\in J$, $n \geq 1,$ $ m_{1},\cdots,m_{n} \geq 0$ and
$\ell_{1},\cdots, \ell_{n} \geq 1$ where $\delta_{G}^m$ denotes
the $m$-th iterated commutator with $G$ and $\delta_{G}^0(L_\ell)=L_\ell$.
\end{theorem}

In our framework the operator $G$ can be written as
\begin{eqnarray*}
G & = & -\left(\frac{\lambda^2+\mu^2}{2}+\mi\left(\zeta^{+}+\zeta^{-}\right)\right) N^2
+ \left(\frac{\lambda^2}{2}+\mi\zeta^{+}\right) r \\
&- &\left(\left(\frac{\lambda^2+\mu^2}{2}+\mi\left(\zeta^{+}+\zeta^{-}\right) \right) r
+ \left(\frac{\lambda^2-\mu^2}{2}+\mi\left(\zeta^{+}-\zeta^{-}\right)\right)\right) N.
\end{eqnarray*}
Note that the self-adjoint operator $N^2$ generates a semigroup
$(e^{z N^2})_{z\in\mathbb{C}-\mathbb{R}_{+}}$ defined on the
complex plane without the positive real half axis which is an
analytic semigroup in the half plane $\{ z\in\mathbb{C}\,\mid\, \Re(z)<0\}$.
Thus, by a change of variable, the operator
\begin{equation}\label{eq:partG}
G_0=-\left(\frac{\lambda^2+\mu^2}{2}+\mi\left(\zeta^{+}+\zeta^{-}\right)\right) N^2
\end{equation}
generates an analytic semigroup in the sector
\[
\Re\left(\left(\frac{\lambda^2+\mu^2}{2}+\mi \left(\zeta^{+}+\zeta^{-}\right)\right)z\right)>0,
\]
which is equivalent to $\Re(z)>0$ for $\zeta^{+}+\zeta^{-}=0$ and
\[
\begin{array}{rl}
\Im(z)< \frac{\lambda^2+\mu^2}{2\left(\zeta^{+}+\zeta^{-}\right)} \Re(z),
& \quad \hbox{if } \zeta^{+}+\zeta^{-} >0, \\
\Im(z)>\frac{\lambda^2+\mu^2}{2\left(\zeta^{+}+\zeta^{-}\right)} \Re(z),
& \quad \hbox{if } \zeta^{+}+\zeta^{-} <0.
\end{array}
\]
In any case the semigroup generated by the operator $G_0$ in (\ref{eq:partG}) is
analytic in the sector
\begin{equation}\label{eq:an-sect-qho}
\left\{ z\in\mathbb{C}\,\mid\, \left|\hbox{\rm arg }(z)\right| <
\arctan\left((\lambda^2+\mu^2)/(2\left|\zeta^{+}+\zeta^{-}\right|\right) \right\}
\end{equation}
with the convention $\arctan(+\infty)=\pi/2$.
Clearly $\Dom(G)=\Dom(G_0)=\Dom(N^2)$ and $G$ is the sum of $G_0$ and an operator
with domain $\Dom(N)=\Dom(G_0^{1/2})$ where $G_0^{1/2}$ is defined as $N$ times
a square root of the complex number $(\lambda^2+\mu^2)/2
+\mi(\zeta^{+}+\zeta^{-})$.
It follows then from  Corollary 2.4 p.81 of \cite{Pazy} that $G$ generates an analytic
semigroup in the sector (\ref{eq:an-sect-qho}).

\medskip

We can now prove the following.

\begin{theorem}\label{th:inst-spread}
For all initial state $\eta$ the support of the
state $\mathcal{T}_{*t}(\eta)$ at any time $t>0$ is the whole of $\h$.
\end{theorem}

\noindent{\bf Proof.}
The assumption (\ref{eq:LGk}) obviously holds because
$\Dom(G^k)=\Dom(N^{2k})$ and $B,B^+$ map $\Dom(N^h)$ in $\Dom(N^{h-1})$ for
all $h$ so that
\[
B^\pm(\Dom(G^k)) \subseteq \Dom(N^{2k-1})
\subseteq\Dom(N^{2k-2})=\Dom(G^{k-1}).
\]

For all state $\omega$ we can write its spectral decomposition
$\omega=\sum_{j \in J}\omega_{j}|\psi_{j}\rangle \langle \psi_{j}|$, for a collection
$(\psi_j)_{j\in J}$ of orthonormal vectors and $\omega_{j}> 0$ for all $j \in J$ and
$J$ non-empty. Since $L_1=\lambda B^{+},L_2=\mu B^{-}$, $\delta_G^{0}(B^\pm)=B^\pm$
and $[B,B^{+}]-r\,\mathbb{1}=2N$, by Theorem \ref{th:HaNaGl}, all vectors
\begin{equation}\label{eq:Nn-supp}
N^nP_t\psi_j = 2^{-n} \left([B,B^{+}]-r\mathbb{1}\right)^nP_t\psi_j
\end{equation}
belong to the support of $\mathcal{T}_{*t}(\omega)$. Write
$P_t\psi_j = \sum_{k\ge 0} z_{jk}(t) e_k$
and let $m_j$ be the minimum $k$ for which $z_{jk}(t)\not=0$. Since the function
$z\to \hbox{\rm e}^{-zN}$ is analytic for $\Re z >0$, and $P_t\psi_j$ belongs to
the support of $\mathcal{T}_{*t}(\omega)$, for all real number $s>0$
\begin{eqnarray*}
\sum_{n\ge 0} \frac{(-s)^n}{n!}(N-m_j\mathbb{1})^n P_t\psi_j
& = & \hbox{\rm e}^{-s(N-m_j\mathbb{1})} P_t\psi_j \\
& = & z_{jm_j}(t) e_{m_j} + \sum_{k>m_j}z_{jk}(t) \hbox{\rm e}^{-s(k-m_j)}e_k
\end{eqnarray*}
belongs to the support of $\mathcal{T}_{*t}(\omega)$ by \eqref{eq:Nn-supp} for
all $s>0$. Taking the limit as $s$ tends to infinity, we conclude that
$e_{m_j}$ belongs to the support of $\mathcal{T}_{*t}(\eta)$. In the same way,
starting from either
\[
B^kP_t\psi_j=(\delta_G^0(B))^k P_t\psi_j \qquad \hbox{or} \qquad B^{+k}P_t\psi_j=(\delta_G^{0}(B^+))^kP_t\psi_j,
\]
we can conclude that
\[
B^k e_{m_j} \qquad \hbox{and} \qquad B^{+k}e_{m_j}
\]
belong to the support of $\mathcal{T}_{*t}(\omega)$ for $t>0$ for all $k$.
As a consequence, the support of $\mathcal{T}_{*t}(\omega)$ is the
whole of $\mathsf{h}$ for all $t>0$.

\section{Spectral gap}\label{sect:gap}

In this section we discuss the spectral gap of the generator of the semigroup of the quadratic
harmonic oscillator. For the purpose we will follow the methods developed in
Refs. \cite{CaFa, CaFaGaQu}.

\subsection{Dirichlet form and spectral gap}

Recall the invariant state $\rho$ in \eqref{eq:inv-state}. Let $L_2(\mathsf h)$ be the space of Hilbert-Schmidt operators on $\mathsf h$ with inner product $\langle x,y\rangle=\tr{x^*y}$. Consider the embedding
 \[
\iota:\mathcal B(\mathsf h)\to L_2(\mathsf h), \quad \iota(x)=\rho^{1/4}x\rho^{1/4}.
\]
Let $T=(T_t)_{t\ge 0}$ be the strongly continuous contraction semigroup on $L_2(\mathsf h)$ defined by
\[
T_t(\iota(x))=\iota(\mathcal T_t(x)).
\]
Let $L$ be the generator of the semigoup $(T_t)_{t\ge 0}$. We can check that
\[
L(\rho^{1/4}x\rho^{1/4})=\rho^{1/4}\mathcal L(x)\rho^{1/4}, \quad \text{for }x\in D(\mathcal L).
\]
\begin{lemma}\label{lem:generator_embedded}
For $x=\sum_{j,k}x_{jk}|e_j\rangle\langle e_k|\in\mathcal M $,
\begin{eqnarray}
 L(x)
& = &  \mi\sum_{j,k \ge 0}(\zeta^{+}(\omega_{j+1}-\omega_{k+1})
+\zeta^{-}(\omega_j-\omega_k))x_{jk} |e_j\rangle\langle e_k| \label{eq-Lindblad-on-M}\\
&&+  \sum_{j,k \ge 0}\Big(\mu\lambda(\omega_k\omega_j)^{1/2}x_{j-1\,k-1}-
{\frac{\mu^2}{2}}(\omega_j+\omega_k)x_{jk} \nonumber \\
&&+  \mu\lambda(\omega_{j+1}\omega_{k+1})^{1/2}x_{j+1\, k+1}
-{\frac{\lambda^2}{2}}(\omega_{j+1}+\omega_{k+1})x_{jk}\Big)|e_j\rangle\langle e_k|.\nonumber
\end{eqnarray}
\end{lemma}
 with the convention $x_{j-1\, k-1}=0$ if $j=0$ or $k=0$.

The Dirichlet form, defined for $\xi\in D(L)$, is the quadratic form $\mathcal E$
\[
\mathcal E(\xi)=-\mathrm{Re}\langle\xi,L(\xi)\rangle.
\]
The spectral gap of the operator $L$ is the nonnegative number
\[
\mathrm{gap}\,L:=\inf\{\mathcal E(\xi):\|\xi\|=1,\,\,\xi\in (\mathrm{Ker}\,L)^\perp\}.
\]
\begin{lemma}\label{lem:Dirichlet_form}
For $\xi\in D(L)$ with $\xi=\sum_{j,k}\xi_{jk}|e_j\rangle\langle e_k|$
\begin{eqnarray*}
\mathcal E(\xi)&=&
\frac12\mu^2\left(\sum_{k\ge 1}\omega_k|\xi_{0k}|^2
+\sum_{j\ge 1}\omega_j|\xi_{j0}|^2\right)\\
&&+\frac12\sum_{j,k\ge 0}\left(\left|\mu\omega_{j+1}^{1/2}\xi_{j+1\,k+1}
-\lambda\omega_{k+1}^{1/2}\xi_{jk}\right|^2+
\left|\mu\omega_{k+1}^{1/2}\xi_{j+1\,k+1}
-\lambda\omega_{j+1}^{1/2}\xi_{jk}\right|^2\right).
\end{eqnarray*}
In particular, $\mathcal E(\xi)=\mathcal E(\xi^*)$, where $\xi^*$ is the adjoint operator of $\xi$.
\end{lemma}

\noindent{\bf Proof. }
By Lemma \ref{lem:generator_embedded} we get
\begin{eqnarray*}
\mathcal E(\xi)&=&-\mathrm{Re\,}\langle\xi,L(\xi)\rangle\\
&=&-\mathrm{Re\,}\sum_{j,k}\Big[\mu\lambda(\omega_j\omega_k)^{1/2}
\overline{\xi_{jk}}\xi_{j-1\,k-1}-\frac12\mu^2(\omega_j+\omega_k)|\xi_{jk}|^2\\
&&\hskip 1 true cm+\mu\lambda(\omega_{j+1}\omega_{k+1})^{1/2}\overline{\xi_{jk}}
\xi_{j+1\,k+1}-\frac12\lambda^2(\omega_{j+1}+\omega_{k+1})|\xi_{jk}|^2\Big]
\end{eqnarray*}
Rearranging the terms we get the desired expression.\hfill $\square$
  \\[1ex]
\begin{proposition}\label{prop:kernel}
If $W=\{\xi\in D(L):\mathcal E(\xi)=0\}$ then
\[
\mathrm{span}\{\rho^{1/2}\}=\iota(\mathrm{Ker\,}\mathcal L)=\mathrm{Ker\,}L=W.
\]
\end{proposition}
\noindent{\bf Proof.}
Since $\mathbb{1}\in \mathrm{Ker}\,\mathcal L$, $\rho^{1/2}=\iota(\mathbb{1})\in \mathrm{Ker}\,L$. It is obvious that $\mathrm{Ker}\,L\subset W$. Suppose $\xi=\sum_{j,k}\xi_{jk}|e_j\rangle\langle e_k|\in W$. By Lemma \ref{lem:Dirichlet_form}
\begin{equation}\label{eq:kernel_condition}
\xi_{j+1\,k+1}=\nu\sqrt{\frac{\omega_{k+1}}{\omega_{j+1}}}\xi_{jk}=\nu\sqrt{\frac{\omega_{j+1}}{\omega_{k+1}}}\xi_{jk}
\end{equation}
and
\[
\xi_{0k}=\xi_{k0}=0,\quad\text{ for }k\ge 1.
\]
Thus $\xi$ must be diagonal and by \eqref{eq:kernel_condition}, $\xi_{k+1\,k+1}=\nu\xi_{kk}$. Hence for $k\ge 0$, $\xi_{kk}=\nu^k\xi_{00}$, i.e., $\xi=\text{const.}\rho^{1/2}$. This completes the proof.\hfill $\square$ \\[1ex]

Like the model of two-photon absorption and emission process discussed in \cite{CaFaGaQu}, there
are invariant subspaces for the process of quadratic open quantum harmonic oscillator.
For $m\in \mathbb Z$, define
\[
\mathcal G_m:=\overline{\mathrm{span}\{|e_k\rangle\langle e_{k+m}|:\,k\ge \mathrm{max}\{0,-m\}\}}.
\]
One can easily check the following properties.
\begin{enumerate}
\item[(1)] $\xi\in \mathcal G_m\Leftrightarrow\xi^*\in \mathcal G_{-m}$.\\[-2ex]
\item[(2)] The linear spaces $\mathcal G_m$ are orthogonal in $L_2(\mathsf h)$,
and
\[
L_2(\mathsf h)=\oplus\{\mathcal G_m:\,m\in \mathbb Z\}.
\] \\[-6ex]
\item[(3)] Each $\mathcal G_m$ is invariant for the generator $L$ and so also for the semigroup $T$.\\[-2ex]
\item[(4)] Each $\mathcal G_m$ is isometrically isomorphic to the space $l^2(\mathbb N)$ of square summable sequences.
\end{enumerate}
By mimicking the proof of \cite[Proposition 4]{CaFaGaQu}, we can show the following.
\begin{proposition}\label{prop:gap_computation}
\[
\mathrm{gap}\,L=\inf_{m\ge 0}A_m,
\]
where $A_m=\begin{cases}\inf\{\mathcal E(\xi):\,\|\xi\|=1,\,\,\xi\in \mathcal G_m\},&\mathrm{for}\,\,m\neq0\\
\inf\{\mathcal E(\xi):\,\|\xi\|=1,\,\,\xi\in \mathcal G_0,\,\,\xi\perp\rho^{1/2}\},&\mathrm{for}\,\,m=0\end{cases}$.
\end{proposition}

We will now study separately off-diagonal and diagonal minima.

\subsection{Off-diagonal minimum problems}
Fix $m>0$. For $\xi\in \mathcal G_m$ we can write for some sequence $y=(y_j)_{j\ge 0}$
in $l^2(\mathbb N)$,
\[
\xi=\sum_jy_j|e_j\rangle\langle e_{j+m}|,
\]
where $y_j=\xi_{j\,j+m}$. Then
\begin{eqnarray}\label{eq:Dirichlet_form_off-diagonal}
\mathcal E(\xi)&=&\frac12\mu^2\omega_m|y_0|^2 +\frac12\sum_j\Big(|\mu\sqrt{\omega_{j+1}}y_{j+1}-\lambda\sqrt{\omega_{j+m+1}}y_j|^2  \nonumber\\
&&\hskip 1 true cm
+|\mu\sqrt{\omega_{j+m+1}}y_{j+1}-\lambda\sqrt{\omega_{j+1}}y_j|^2\Big).
\end{eqnarray}

\begin{proposition}\label{prop:off-diagonal_computation}
For any $m\ge 1$,

\begin{equation}\label{eq:off-diagonal_bound}
A_m\ge \frac{\mu^2 m}{2}\left( (m+r-1) + \nu^2(m-r+1)\right)
\end{equation}
The lower bound is attained by $\mathcal{E}(\xi)$ for
\[
y_j^2=\nu^{2(j+1)}\binom{j+m}{m}\frac{\Gamma(m+r)\Gamma(j+r)}{\Gamma(r)\Gamma(j+m+r)}y_0^2
\]
and $y_0\ge 0$ such that $\Vert \xi\Vert^2=1$.
\end{proposition}

\noindent{\bf Proof.}
We fix $m\ge 1$ and $\xi\in \mathcal G_m$. From the formula \eqref{eq:Dirichlet_form_off-diagonal}, without loss of generality we may assume $y_j\ge 0$ for all $j\ge 0$ because $|a-b|\ge ||a|-|b||$ for any complex numbers $a$ and $b$. Then we can rewrite
\begin{eqnarray}\label{eq:positive_components}
\mathcal E(\xi)&=&\frac12\mu^2\omega_my_0^2+\frac12\sum_{j\ge 0}\Big((\omega_{j+1}+\omega_{j+m+1})(\mu^2y_{j+1}^2
+\lambda^2y_j^2)\nonumber\\
&&\hskip 1true cm -4\mu\lambda\sqrt{\omega_{j+1}\omega_{j+m+1}}y_{j+1}y_j\Big).
\end{eqnarray}
Since
\begin{equation}\label{eq:Schwarz-weight}
2\lambda\mu\, y_{j+1}y_{j}\le \mu^2 \theta_j y_{j+1}^2 + \frac{\lambda^2}{\theta_j}y_j^2
\end{equation}
where $\theta_j>0$ is such that
\[
\omega_{j+m+1}+\omega_{j+1}-2\theta_j\sqrt{\omega_{j+m+1}\omega_{j+1}} = \omega_m
\]
i.e.
\[
\theta_{j}=\frac{\omega_{j+m+1}+\omega_{j+1}-\omega_m}{2\sqrt{\omega_{j+m+1}\omega_{j+1}}}.
\]
In this way, we get the inequality
\begin{eqnarray*}
\mathcal E(\xi)&\ge&\frac{\mu^2}{2}\omega_my_0^2+\frac{\mu^2}{2}\sum_{j\ge 0}\omega_m y_{j+1}^2
+\frac{\lambda^2}{2}\sum_{j\ge 0}
\left(\omega_{j+m+1}+\omega_{j+1}-\frac{2\sqrt{\omega_{j+m+1}\omega_{j+1}}}{\theta_j}\right)y_j^2\\
&= & \frac{\mu^2}{2}\omega_m \Vert y \Vert^2
+\frac{\lambda^2}{2}\sum_{j\ge 0}
\left(\omega_{j+m+1}+\omega_{j+1}
-\frac{4\omega_{j+m+1}\omega_{j+1}}{\omega_{j+m+1}+\omega_{j+1}-\omega_m}\right)y_j^2\\
& = & \frac{\mu^2}{2}\omega_m \Vert y \Vert^2
+\frac{\lambda^2}{2}\sum_{j\ge 0}
\frac{m(m-r+1)(j+1)(j+m+r)}{(1+j)(j+m+r)}y_j^2\\
& = & \frac{\mu^2}{2}\omega_m \Vert y \Vert^2
+\frac{\lambda^2}{2}\sum_{j\ge 0}m(m-r+1)y_j^2.
\end{eqnarray*}
It follows that 
\begin{eqnarray*}
A_m\ge \frac{\mu^2\omega_m}{2} +\frac{\lambda^2m(m-r+1)}{2}
=\frac{\mu^2 m}{2}\left((m+r-1) + \nu^2(m-r+1)\right)
\end{eqnarray*}
The above lower bound for the Dirichlet form is attained if and only if the
Schwarz inequalities (\ref{eq:Schwarz-weight}) are equalities namely
$\mu \theta_j^{1/2} y_{j+1} = \lambda \theta_j^{-1/2} y_j$ so that
\[
y_{j+1} = \frac{\nu}{\theta_j} y_j
\]
for all $j\ge 1$. Iterating we find
\[
y_{j+1} = \frac{\nu^{j+1}}{\theta_j\cdots\theta_0} y_0
= \nu^{j+1} y_0 \prod_{i=0}^{j}
\frac{2\sqrt{\omega_{i+m+1}\omega_{i+1}}}{\omega_{i+m+1}+\omega_{i+1}-\omega_m}.
\]
Since
\[
\lim_{j\to\infty}\frac{y_{j+1}^2}{y_j^2}
= \nu^2 \lim_{j\to\infty}\frac{4\omega_{j+m+1}\omega_{j+1}}{(\omega_{j+m+1}+\omega_{j+1}-\omega_m)^2}
=\nu^2 < 1,
\]
we find $\sum_{j\ge 0} y_j^2 <\infty$ and so the lower bound is a minimum. \\
Minimizers can be written in an explicit form. First note that
\[
y_{j+1}^2 =\nu^2 \frac{4(j+m+1)(j+m+r)(j+1)(j+r)}{4(j+1)^2(j+m+r)^2}y_j^2
=\nu^2 \frac{(j+m+1)(j+r)}{(j+m+r)(j+1)}y_j^2.
\]
Iterating
\begin{eqnarray*}
y_{j+1}^2 & = & \nu^{2(j+1)}\frac{(j+m+1)\cdots (m+1)}{(j+m+r)\cdots (m+r)}
\frac{(j+r)\cdots r}{(j+1)!}y_0^2 \\
& = & \nu^{2(j+1)}\frac{(j+m+1)!}{m!(j+1)!} \frac{(j+r)\cdots r}{(j+m+r)\cdots (j+1+r)} y_0^2 \\
& = & \nu^{2(j+1)}\binom{j+m+1}{m}\frac{\Gamma(m+r)\Gamma(j+r+1)}{\Gamma(r)\Gamma(j+m+r+1)}y_0^2.
\end{eqnarray*}
\hfill $\square$

\medskip
\noindent The function $m\to (m+r-1) + \nu^2(m-r+1)$ is a positive and increasing function
of $m$ for $m\ge 1$ and so we have
\begin{theorem}\label{th:off-diag-min}
The off-diagonal gap is
\[
\inf_{m\ge 1}\frac{\mu^2 m}{2} \left((m+r-1) + \nu^2(m-r+1)\right)
=\frac{\mu^2}{2}\left(2\nu^2 +(1-\nu^2)r\right).
\]
\end{theorem}

\subsection{Diagonal minima}
For any $f$ in $l^2(\mathbb N)$, let us denote by $M_f\in \mathcal B(\mathsf h)$ the multiplication operator by $f$, $M_f=\sum_jf_j|e_j\rangle\langle e_j|$. Then we get
\[
\mathcal L(M_f)=M_{\mathcal Af}\quad \text{for }f\in D(\mathcal A):=\{f\in l^\infty:\,M_f\in D(\mathcal L)\}.
\]
The (formal) explicit expression for $\mathcal A$ is given by
\[
\mathcal Af=\sum_j\left(\lambda^2\omega_{j+1}(f_{j+1}-f_j)
+\mu^2\omega_j(f_{j-1}-f_j)\right)e_j.
\]

The invariant measure for this classical birth and death process is
\[
\pi=(\pi_u)_{u\ge 0}\text{ with }\pi_u=(1-\nu^2)\nu^{2u}.
\]
From now on, whenever there is no confusion we write simply $f$ for $M_f\in \mathcal B(\mathsf h)$. By Lemma \ref{lem:Dirichlet_form} we can compute
\[
\mathcal E(\pi^{1/2}f)=\sum_{j}\lambda^2\omega_{j+1}|f_{j+1}-f_j|^2\pi_j.
\]
\begin{proposition}\label{prop:diagonal_gap_bound}
For any positive sequence $(a_n)_n$, define the (strictly positive) constant
\[
B(\nu):=\sup_{u\ge 0}\frac{(\mu^2(1-\nu^2))^{-1}}{a_u\omega_{u+1}}\left\{\sum_{v\le u}(1-\nu^{2(v+1)})a_v+(\nu^{-2(u+1)}-1)\sum_{v>u}\nu^{2(v+1)}a_v\right\}.
\]
Then $\mathrm{gap}(\mathcal A)\ge B(\nu)^{-1}$.
\end{proposition}

\noindent{\bf Proof.}
We follow the proof of \cite[Proposition 7]{CaFaGaQu} with a change $(2u+1)(2u+2)$
by $\omega_{u+1}$, which amounts to consider the birth rate $b_j=\lambda^2\omega_{j+1}$
instead of $b_j=\lambda^2(2j+1)(2j+2)$.
\hfill $\square$\\[1ex]
The following was proven in \cite[Lemma 8]{CaFaGaQu}.
\begin{lemma}\label{lem:bound_computation}
Take a positive summable sequence $(a_n)_{n\ge 0}$ and define the positive decreasing tail sequence $(A_k)_{k\ge 0}$ by $A_k=\sum_{n\ge k}a_k$. Then
\[
B(\nu)=\sup_{u\ge 0}\frac1{\mu^2a_u\omega_{u+1}}\sum_{k\ge 0}\nu^{2k}(A_k-A_{k+u+1}).
\]
\end{lemma}
Thus the computation of the spectral gap relies on how we choose the sequence $(a_n)_n$.
Here we propose the following choice. (cf. \cite[Lemma 9]{CaFaGaQu})
\begin{lemma}\label{lem:sequence}
Let $a_n=\frac1{(n+r+1)(n+r)}$. Then the following properties hold.
\begin{enumerate}
\item[(i)] $A_k=\sum_{n\ge k}a_n=\frac1{k+r}$.\\[-3ex]
\item[(ii)] $\sum_{k\ge 0}\nu^{2k}a_k=2\nu^{-2(r+1)}\int_0^\nu\frac{s^{2r-1}}{1-s^2}(\nu^2-s^2)ds$.\\[-3ex]
\item[(iii)] $\sum_{k\ge 0}\nu^{2k}A_k=2\nu^{-2r}\int_0^\nu\frac{s^{2r-1}}{1-s^2}ds$.
\end{enumerate}
\end{lemma}

\noindent{\bf Proof.}
The item (i) is trivial. To prove (ii), we write
\[
\sum_{k\ge 0}\nu^{2k}a_k=\sum_{k\ge 0}\nu^{2k}\left(\frac1{k+r}-\frac1{k+r+1}\right).
\]
Let
\[
a(\nu):=\sum_{k\ge 0}\nu^{2k}\frac1{k+r}.
\]
Then, $b(\nu):=a(\nu)\nu^{2r}$ satisfies
\[
b'(\nu)=2\frac{\nu^{2r-1}}{1-\nu^2}.
\]
Since $b(0)=0$, we get
\[
b(\nu)=2\int_0^\nu\frac{s^{2r-1}}{1-s^2}ds.
\]
From this we easily get the desired expression. For (iii), we see by (i) that the value we are looking for is $a(\nu)$, which we computed in the above.
\hfill $\square$

\medskip
We can now find a lower bound for the diagonal minimum of the
Dirichlet form on vectors $\xi$ orthogonal to $\rho^{1/2}$.
\begin{theorem}\label{thm:diagonal_gap_bound}
\[
\mathrm{gap}(\mathcal A)\ge \mu^2\left(\sum_{k\ge 0}\nu^{2k}\frac1{k+r}\right)^{-1}=  \mu^2\nu^{2r}\left(2\int_0^\nu\frac{s^{2r-1}}{1-s^2}ds\right)^{-1}.
\]
\end{theorem}

\noindent{\bf Proof.}
We choose the sequence $(a_n)_n$ as in Lemma \ref{lem:sequence}. Let us define a function
$V:\mathbb N\to \mathbb R$ by
\begin{eqnarray*}
V(u)&:=&\frac1{a_u\omega_{u+1}}\sum_{k\ge 0}\nu^{2k}(A_k-A_{k+u+1})\\
&=&\frac{u+r+1}{u+1}\sum_{k\ge 0}\nu^{2k}\left(\frac1{k+r}-\frac1{k+u+1+r}\right).
\end{eqnarray*}
We can see that the function $V(u)$ is increasing. In fact, regarding $V(u)$ as a function on the interval $[0,\infty)$, we differentiate it. With a little computation we see that
\[
V'(u)=\sum_{k\ge 0}\nu^{2k}\frac{k}{(k+u+1+r)^2(k+r)}>0.
\]
Therefore we get by using Lemma \ref{lem:sequence} (iii)
\[
\mu^2B(\nu) = \sup_{u\ge 0}V(u)
= \sum_{k\ge 0}\nu^{2k}\frac1{k+r}
= 2\nu^{-2r}\int_0^\nu\frac{s^{2r-1}}{1-s^2}ds.
\]
The result now follows by Proposition \ref{prop:diagonal_gap_bound}.
\hfill $\square$

\medskip
{\bf Remark.} The lower bound of theorem \ref{thm:diagonal_gap_bound}
can be written in a closed form by introducing the Lerch function
$\Phi:\left\{\,z\in\mathbb{C}\,\mid\, |z|<1\,\right\}\times \mathbb{N}\times
]0,+\infty[\to \mathbb{C}$
\[
\Phi(z,n,r) = \sum_{k\ge 0}\frac{z^{k}}{(k+r)^n}.
\]
By Theorem \ref{thm:diagonal_gap_bound}, we have $\mathrm{gap}(\mathcal A)\ge\mu^2
\left(\Phi(\nu^2,1,r)\right)^{-1}$. Moreover, by comparing the off-diagonal
explicit minimum and the diagonal lower bound we find the following.

\begin{theorem}\label{thm:gap}
For all $\nu,r$ such that $\left(2\nu^2+r(1-\nu^2)\right)\Phi(\nu^2,1,r)\le 2$
we have
\begin{equation}\label{eq:gap-sharp}
\mathrm{gap}\,L=\frac{\mu^2}2\left(2\nu^2+r(1-\nu^2)\right).
\end{equation}
In particular, if $r\ge 2\nu^2/(1-\nu^2)$ the above identity holds.
\end{theorem}

\noindent{\bf Proof.}
The first claim follows immediately by comparing the
diagonal lower bound of Theorem \ref{thm:diagonal_gap_bound} and the
off-diagonal minimum of Theorem \ref{th:off-diag-min}.
By the elementary inequality,
\[
\sum_{k=0}^\infty\frac{\nu^{2k}}{k+r}=\frac1r\sum_{k=0}^\infty\frac{r\nu^{2k}}{k+r}\le \frac{1}{r}\sum_{k=0}^\infty\nu^{2k}=\left(r(1-\nu^2)\right)^{-1}
\]
we have $\Phi(\nu^2,1,r)\le\left(r(1-\nu^2)\right)^{-1}$ and so
the identity \eqref{eq:gap-sharp} holds, in particular,
if $\left(r(1-\nu^2)\right)^{-1}\left(2\nu^2+r(1-\nu^2)\right)\le 2$,
i.e. $r\ge 2\nu^2/(1-\nu^2)$.
\hfill $\square$

\medskip
The graph in Figure 1 shows the values of $\nu,r$
for which the identity $\left(2\nu^2+r(1-\nu^2)\right)\Phi(\nu^2,1,r)=2$ holds. Clearly,
for pairs $(\nu,r)$ lying above (or on) the graph the spectral gap is given by
\eqref{eq:gap-sharp}.
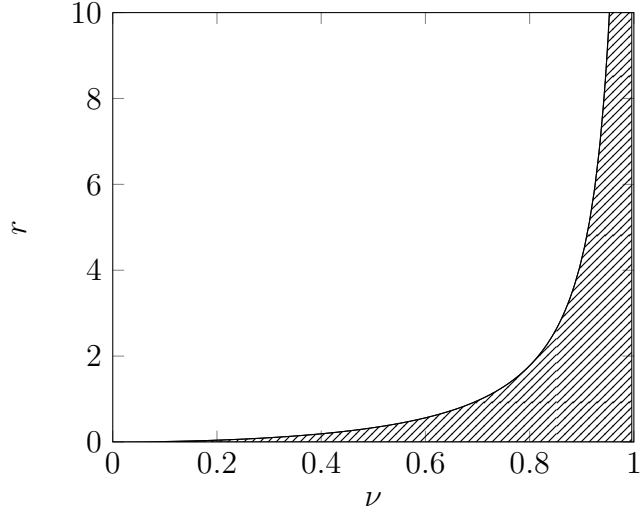
\begin{figure}[ht!]
\centering
\begin{tikzpicture}
    \begin{axis}[xmin=0,xmax=1,xlabel={$\nu$},ylabel={$r$},ymin=0,ymax=10]
    \addplot[color=black,domain=0:.996,samples=100] {x^2/(1-x^2)};
    \addplot[color=black,fill=black, pattern=north east lines,  domain=0.1:.996,samples=100]
    {x^2/(1-x^2)} \closedcycle;
    \end{axis}
\end{tikzpicture}
\caption{gap$(L)=\mu^2(2\nu^2+r(1-\nu^2))/2$, off-diagonal minimum,
 for $(\nu,r)$ above the curve.}
\end{figure}
The exact value of the spectral gap, for pairs $(\nu,r)$ lying below the
graph, could be also the diagonal minimum whose exact value is not known and
we are unable to compute.
This will be clear studying upper bounds for the diagonal spectral gap.

\subsection{Upper bound}
In this section we discuss the upper bound of the diagonal spectral gap.
By definition, any value $\mathcal E(\xi)$ with $\|\xi\|=1$, with $\xi$ orthogonal
to $\rho^{1/2}$ is an upper bound for the gap. Thus to get a good upper bound we
need to choose a vector $\xi$ cleverly.

Looking at the explicit form of the off-diagonal minimizer that we get for $m=1$,
we consider the first order polynomial $f_j=j-c$ where $c$ is a constant chosen in such
a way that $j\to\pi_j^{1/2}f_j$ is orthogonal to $\pi_j^{1/2}$ i.e. $\sum_{j}\pi_j(j-c)=0$.
It is worth noticing here that this choice yields the desired minimizer of the
Dirichlet form on vectors orthogonal to $\rho^{1/2}$ for the usual harmonic oscillator
(see \cite{CaFa}). Using the identities
\[
(1-\nu^2)\sum_{k\ge 0} k \nu^{2k} =\frac{\nu^2}{1-\nu^2}, \qquad
(1-\nu^2)\sum_{k\ge 0} k^2 \nu^{2k} =\frac{\nu^2(1+\nu^2)}{(1-\nu^2)^2},
\]
one computes $c=\nu^2/(1-\nu^2)$. Then, considering $\xi_j=\pi^{1/2}_j(j-c)$,
one finds
\begin{eqnarray*}
\Vert \xi \Vert^2 & = & (1-\nu^2)\sum_{k\ge 0} (k-c)^2 \nu^{2k}
=\frac{\nu^2}{(1-\nu^2)^2} \\
\mathcal{E}(\xi)&=&\lambda^2\sum_{j\ge 0}(f_{j+1}-f_j)^2(j+1)(j+r)\pi_j
= \lambda^2\frac{2\nu^2 + r(1-\nu^2)}{(1-\nu^2)^2}
\end{eqnarray*}
and so
\[
\frac{\mathcal{E}(\xi)}{\Vert \xi \Vert^2}
= \mu^2(2\nu^2 + r(1-\nu^2)).
\]
As a consequence, one has the explicit upper bound
\[
\mathrm{gap}(L) \le \mu^2(2\nu^2 + r(1-\nu^2))
\]
which is twice the spectral gap in good cases by Theorem \ref{thm:gap}.

\bigskip

In order to show that the spectral gap converges to $0$ as $r\to 0$, we find
another upper bound computing the value of the Dirichlet form for another vector
$\xi$ suggested by our choice of the sequence $(A_k)_k$ in
Lemma \ref{lem:sequence}.
\begin{theorem}\label{thm:gap_upper bound}
\[
\mathrm{gap}\,L\le \lambda^2\frac{\sum_j\frac{j+1}{(j+r+1)^2(j+r)}\nu^{2j}}
{\sum_j\left(\frac1{j+r}-c(\nu,r)\right)^2\nu^{2j}},
\]
where $c(\nu,r)=(1-\nu^2)\sum_{j=0}^\infty\frac{\nu^{2j}}{j+r}=(1-\nu^2)\Phi(\nu^2,1,r)$.
In particular, for all $\nu>0$ fixed, $\mathrm{gap}\,L$ tends to $0$ as $r\to 0$.
\end{theorem}

\noindent{\bf Proof.}
We consider $A_k= (k+r)^{-1}$ as in Lemma \ref{lem:sequence}, and define
$f_k=A_k-c=(k+r)^{-1}-c(\nu,r)$, where the constant $c$ is such that $\sum_j\pi_jf_j=0$.
We compute
\begin{eqnarray*}
\mathcal{E}(\pi^{1/2}h)&=&\sum_{j\ge 0}\lambda^2\omega_{j+1}|f_{j+1}-f_j|^2\pi_j\\
&=&(1-\nu^2)\lambda^2\sum_{j\ge 0}\frac{j+1}{(j+r+1)^2(j+r)}\nu^{2j}.
\end{eqnarray*}
On the other hand,
$\|f\|_{\pi}^2=(1-\nu^2)\sum_{j\ge 0}\left((j+r)^{-1}-c(\nu,r)\right)^2\nu^{2j}.$
Therefore,
\begin{equation}\label{eq:upper_bound_diagonal}
\mathcal{E}(\xi)/\|\xi\|^2\le \lambda^2\frac{\sum_j\frac{j+1}{(j+r+1)^2(j+r)}\nu^{2j}}
{\sum_j\left(\frac1{j+r}-c\right)^2\nu^{2j}}
\end{equation}
We now check that the above upper bound tends to $0$ as $r\to 0$. To this end note that
we can write the denominator as
\begin{eqnarray*}
& & \sum_{j\ge 0}\frac{\nu^{2j}}{(j+r)^2}-(1-\nu^2)\left(\sum_{j\ge 0}\frac{\nu^{2j}}{j+r}\right)^2 \\
& = & \frac{1}{r^2} \left(1+\sum_{j\ge 1}\frac{r^2\nu^{2j}}{(j+r)^2}\right)
+(1-\nu^2) \left(\frac{1}{r}+\sum_{j\ge 1}\frac{\nu^{2j}}{j+r}\right)^2
\end{eqnarray*}
It follows that, if we multiply the denominator by $r^2$ in the limit as $r\to 0$
we get $1-(1-\nu^2)=\nu^2$. On the other hand, if we multiply also the numerator by $r^2$
we get
\[
\lim_{r\to 0}\sum_{j\ge 0}\frac{r^2(j+1)\nu^{2j}}{(j+r+1)^2(j+r)}
\le \lim_{r\to 0}\sum_{j\ge 0}\frac{r\nu^{2j}}{(j+r+1)}
\le \lim_{r\to 0}r\sum_{j\ge 0}\frac{\nu^{2j}}{(j+1)} = 0.
\]
This completes the proof. \hfill $\square$

\bigskip
{\bf Conclusion.}  Theorem \ref{thm:gap_upper bound} shows that,
for $r$ near $0$,  $\mathrm{gap}\,L$ is the diagonal gap whose exact value is not known.
Moreover, it tends to $0$ as $r\to 0$. The exact result of Theorem \ref{thm:gap} holds in
the white region above the graph in Figure 1. It is worth noticing here that the range
of values of $r$ for which our exact result holds becomes closer and closer to the
half-line $(0,+\infty)$ as $\nu^2=\hbox{\rm e}^{-s\beta}$ goes to $0$, namely
the inverse temperature $\beta$ goes to $+\infty$.
This confirms the intuition that quantum (off-diagonal) effects prevail over the classical
(diagonal) ones when the temperature is small.

\section*{Acknowledgement}
A. Dhahri acknowledges support by {Fondo professori stranieri} DHG9VARI01 Politecnico di Milano.
The research by H. J. Yoo was supported by Basic Science Research Program through the National
Research Foundation of Korea (NRF) funded by the Ministry of Education (NRF-2016R1D1A1B03936006).

\end{document}